\documentclass[aps,prb,twocolumn,amsmath,amssymb,superscriptaddress,longbibliography]{revtex4-2}

\usepackage{CJK}
\usepackage{graphicx} 
\usepackage{bm}
\usepackage[pdfstartview=FitH, CJKbookmarks=true, bookmarksnumbered=true, bookmarksopen=true, colorlinks=true, pdfborder=0 0 1, citecolor=blue, linkcolor=blue, linktocpage=true] {hyperref}
\usepackage{epstopdf} 

\begin{document}

\title{Real-space representation of winding number for one-dimensional chiral-symmetric topological insulator}

\author{Ling Lin}
\affiliation{Guangdong Provincial Key Laboratory of Quantum Metrology and Sensing $\&$ School of Physics and Astronomy, Sun Yat-Sen University (Zhuhai Campus), Zhuhai 519082, China}
\affiliation{State Key Laboratory of Optoelectronic Materials and Technologies, Sun Yat-Sen University (Guangzhou Campus), Guangzhou 510275, China}

\author{Yongguan Ke}
\email{keyg@mail2.sysu.edu.cn}
\affiliation{Guangdong Provincial Key Laboratory of Quantum Metrology and Sensing $\&$ School of Physics and Astronomy, Sun Yat-Sen University (Zhuhai Campus), Zhuhai 519082, China}

\author{Chaohong Lee}
\email{lichaoh2@mail.sysu.edu.cn}
\affiliation{Guangdong Provincial Key Laboratory of Quantum Metrology and Sensing $\&$ School of Physics and Astronomy, Sun Yat-Sen University (Zhuhai Campus), Zhuhai 519082, China}
\affiliation{State Key Laboratory of Optoelectronic Materials and Technologies, Sun Yat-Sen University (Guangzhou Campus), Guangzhou 510275, China}
\date{\today}

%

\begin{abstract}
The winding number has been widely used as an invariant for diagnosing topological phases in one-dimensional chiral-symmetric systems.
We put forward a real-space representation for the winding number.
Remarkably, our method reproduces an exactly quantized winding number even in the presence of disorders that break translation symmetry but preserve chiral symmetry.
We prove that our real-space representation of the winding number, the winding number defined through the twisted boundary condition, and the real-space winding number derived previously in [Phys. Rev. Lett. 113, 046802 (2014)], are equivalent in the thermodynamic limit at half filling.
Our method also works for the case of filling less than one half, where the winding number is not necessarily quantized.
Around the disorder-induced topological phase transition, the real-space winding number has large fluctuations for different disordered samples, however, its average over an ensemble of disorder samples may well identify the topological phase transition.
Besides, we show that our real-space winding number can be expressed as a Bott index, which has been used to represent the Chern number for two-dimensional systems.
\end{abstract}

\maketitle

\section{Introduction}
Topological states have attracted a tremendous amount of studies in various systems involving electrons, cold atoms, and photons, etc.
Most of the non-interacting topological states can be successfully explained by topological band theory \citep{kane2013topological} based on perfect translation symmetry \citep{PhysRevLett.61.2015,PhysRevLett.95.226801,bernevig2006quantum,PhysRevLett.96.106802}.
The first well-known example is the integer quantum Hall effect \citep{PhysRevLett.45.494}, in which the quantized Hall conductivity is related to a topological TKNN invariant \citep{PhysRevLett.49.405} defined with Bloch wavefunctions of filling bands.
However, in reality, there always exists disorder that breaks translation symmetry and hence the consequent Bloch wavefunctions. 
It becomes problematic for the calculation of topological invariants with Bloch wavefunctions.
Numerous topological states immune to disorder \citep{bernevig2013topological,PhysRevLett.98.106803,PhysRevLett.106.166802,PhysRevLett.95.136602,PhysRevB.73.045322,PhysRevLett.96.106401,PhysRevB.74.045125,PhysRevLett.110.236803} indicate that the lack of Bloch wavefunctions should not be a hindrance for defining topological invariants.
To circumvent the absence of translation symmetry, one may consider a real-space representation of the topological invariants.
One of the well-known examples is the real-space representation of the Chern number.
The construction of real-space representation of the Chern number is through transforming the momentum-space formula to the real-space one \citep{PhysRevLett.105.115501,Prodan_2011,PhysRevB.84.241106}, or considering a Bott index \citep{Loring_2010,yi2013coupling}.
The real-space representation of the Chern number has been widely used in studying disorder effects in two-dimensional topological systems \citep{PhysRevX.6.011016,PhysRevLett.121.026801,loring2015k,PhysRevB.86.155445}.
Another example is that the polarization in one dimension can be calculated in the real space as well via the projected position operator approach \citep{PhysRevB.26.4269,niu1991theory,RevModPhys.84.1419,PhysRevB.96.245115}.
Recently, a real-space representation of the winding number is proposed for one-dimensional (1D) topological insulators with chiral symmetry \citep{PhysRevLett.113.046802}, in which the momentum-space formula of winding number is transformed to a real-space formula.
Provided that disorder does not break the chiral symmetry, such a real-space representation of the winding number has been proved to be valid and widely used in exploring the topological Anderson insulator \citep{PhysRevB.89.224203,PhysRevB.100.205302,PhysRevB.101.224201,PhysRevResearch.2.043233},  and particularly, for detection of the winding number in experiment \citep{meier2018observation,Maffei2018}.
On the other hand, we note the momentum-space winding number for 1D systems can be written as the `skew' polarization \citep{PhysRevLett.113.046802,RevModPhys.88.035005}, that is, the difference of polarizations (Berry phases) between two sublattices.
As the usual polarization for 1D lattices can be obtained via the projected position operator in real space, \emph{can we derive the real-space winding number in views of the skew polarization}?
In this paper, we propose a real-space representation of the winding number for 1D chiral-symmetric topological insulators.
We use the singular value decomposition (SVD) method to derive a formula for calculating the difference of polarization between two sublattices.
Our formula can be written in form of the Bott index \citep{exel1991invariants,hastings2010almost}, which produces a strictly quantized winding number.
We prove that our formula is exactly equivalent to the momentum-space winding number in the presence of translation symmetry.
We also prove that our real-space representation of the winding number, the winding number defined through the twisted boundary condition (TBC), and the real-space winding number derived previously in Ref.~\citep{PhysRevLett.113.046802}, are equivalent in the thermodynamic limit at half filling.
Provided the chiral symmetry is preserved, our formula is self-averaging and satisfies the bulk-edge correspondence in the presence of disorder.
We have verified numerically that our results are in agreement with the previous works \citep{PhysRevLett.113.046802,PhysRevB.89.224203} on the disordered model after averaging over many random realizations.
However, our method gives exactly quantization of winding number in each realization of disorder, in stark contrast to previous methods \citep{PhysRevLett.113.046802,PhysRevB.89.224203}.
%
%
Away from topological transition points, our method has advantages over the previous method \citep{PhysRevLett.113.046802,PhysRevB.89.224203} for higher accuracy and less fluctuation.
Furthermore, we show that our formula can work for the case of filling less than one half.
Also, we find our real-space representation of the winding number in one dimension can be written as the Bott index \citep{Loring_2010,loring2015k}, which was used to define the real-space representation of the Chern number in two dimensions.
The rest of this paper is organized as follows:
In Sec.~\ref{sec:bulk_polarization_and_projected_position_operator}, we review the projected position operator approach, and show that it is related to the Wilson loop.
In Sec.~\ref{sec:Winding_Number}, we employ SVD for chiral-symmetric Hamiltonian to obtain the flattened Hamiltonian, and then construct a real-space representation of the winding number.
We will prove the equivalence of our real-space representation of winding number and the twisted-boundary winding number.
In Sec.~\ref{sec:application_BDI_model}, we apply our arguments to a 1D toy model belonging to BDI classe.
Finally, in Sec.~\ref{sec:discussion}, we make a summary and discussion.

\section{Bulk polarization and projected position operator}

\label{sec:bulk_polarization_and_projected_position_operator}
In this section, we give a brief review on calculating the bulk polarization in the 1D system via the projected position operator.
Firstly, we consider a finite 1D lattice with $L$ cells under periodic boundary conditions.
When translational symmetry is present, quasi-momentum $k$ becomes a good quantum number, and the Hamiltonian has a block-diagonal structure.
The eigenstates are Bloch waves labeled by quasi-momentum $k$.
The basis of momentum space is the Fourier transform of real space basis.
In the following context, we use $|l,\alpha\rangle$ to refer to the state that a particle is located at the $\alpha$-th sublattice (orbital) of the $l$th cell.
Thus, the basis of momentum space reads
\begin{equation}
|k,\alpha \rangle  = \frac{1}{\sqrt{L}}\sum\limits_{l = 1}^L {{e^{ikl}}|l,\alpha \rangle },
\end{equation}
and the Bloch waves can be written as the linear combinations of momentum basis
\begin{equation}
|{\psi _{k,n}}\rangle  = \sum\limits_\alpha  {u_{k,\alpha }^n|k,\alpha \rangle },
\end{equation}
where $n$ is the index of the band.
The bulk polarization can be calculated through the following formula \citep{PhysRevB.96.245115}
\begin{equation}
\label{eqn:Wilson_loop}
p =   \frac{1}{{2\pi i}}\log \det {{\cal W}_{k + 2\pi  \leftarrow k}},
\end{equation}
where ${{\cal W}_{k + 2\pi  \leftarrow k}}$ is the so-called Wilson loop
\begin{equation}
{{\cal W}_{k + 2\pi  \leftarrow k}} = {F_{k + 2\pi  - \delta k}}{F_{k + 2\pi  - 2\delta k}}...{F_k}.
\end{equation}
The matrix element of $F_k$ is ${\left( {{F_k}} \right)_{m,n}} = \langle u_{k + \delta k}^m|u_k^n\rangle $, in which $m,n$ are the indices of occupied bands.
We will also use this notation in the following context.
Next, we show the Wilson loop can be derived from the projected position operator ${P_{{\rm{occ}}}}\mathcal{X}{P_{{\rm{occ}}}} $, where ${P_{{\rm{occ}}}} = \sum_{n = 1}^{{n_{{\rm{occ}}}}} {\sum_k {|{\psi _{n,k}}\rangle \langle {\psi _{n,k}}|} } $ is the projector onto occupied bands,  and $|{\psi _{n,k}}\rangle$ is the eigenstate of system at the $n$th band with quasi-momentum $k$.
A quantum-mechanical position operator \citep{PhysRevLett.80.1800} $\hat{\mathcal{X}}=\text{exp}( i\delta k\hat{X} )$, in which $\hat{X}=\sum\nolimits_{x}{x{{{\hat{n}}}_{x}}}$ is the general position operator, is introduced for a lattice with periodic boundary condition.
Here, $\delta k=2\pi/L$ is the increment of discrete quasi-momentum $k$.
Note that operator $\mathcal X$ is actually the translation operator for quasi-momentum $k$: $\mathcal{X}|k,\alpha\rangle = |k+\delta k,\alpha\rangle$,
By expanding the expression of projected position operator, we have
\begin{eqnarray}
\label{eqn:PXP_expand}
{P_{{\rm{occ}}}} \mathcal{X}{P_{{\rm{occ}}}} &=& \sum\limits_{n,n' = 1}^{{n_{{\rm{occ}}}}} {\sum\limits_{k,k'} {|{\psi _{n',k'}}\rangle \langle {\psi _{n',k'}}|\mathcal{X}|{\psi _{n,k}}\rangle \langle {\psi _{n,k}}|} }  \nonumber \\
&=& \sum\limits_{n,n' = 1}^{{n_{{\rm{occ}}}}}  \sum\limits_{k}{\langle u_{k + \delta k}^{n'}|u_k^n\rangle |{\psi _{n',k + \delta k}}\rangle \langle {\psi _{n,k}}|}
\end{eqnarray}
where we have used the relation \citep{PhysRevB.56.12847,PhysRevX.6.021008}
\begin{eqnarray}
\langle {\psi _{n',k'}}|\mathcal{X}|{\psi _{n,k}}\rangle &=& \sum\limits_{\alpha ,\alpha '} {{{\left( {u_{k',\alpha '}^{n'}} \right)}^*}u_{k,\alpha }^n\langle k',\alpha '|\mathcal{X}|k,\alpha \rangle } \nonumber \\
&=& \sum\limits_{\alpha ,\alpha '} {{{\left( {u_{k',\alpha '}^{n'}} \right)}^*}u_{k,\alpha }^n\langle k',\alpha '|k+\delta k,\alpha \rangle } \nonumber \\
&=& \sum\limits_{\alpha ,\alpha '} {{\delta _{k',k + \delta k}}{\delta _{\alpha ,\alpha '}}{{\left( {u_{k',\alpha '}^{n'}} \right)}^*}u_{k,\alpha }^n} \nonumber \\
&=& {\delta _{k',k + \delta k}}\langle u_{k'}^{n'}|u_k^n\rangle .
\end{eqnarray}
Then, we seek the eigenvalues of the projected position operator \eqref{eqn:PXP_expand}.
Assuming its eigenstates are the superposition of occupied Bloch states $|\Psi \rangle  = \sum\nolimits_{k,n} {{\Psi _{n,k}}|{\psi _{n,k}}\rangle } $, the eigenvalue problem reads
\begin{eqnarray}
\label{eqn:eigen_problem_PXP}
{P_{{\rm{occ}}}}\mathcal{X}{P_{{\rm{occ}}}}|\Psi \rangle  = \lambda |\Psi \rangle.
\end{eqnarray}
Combining Eq.~\eqref{eqn:PXP_expand} and Eq.~\eqref{eqn:eigen_problem_PXP}, we obtain the following iterative relation
\begin{equation}
\sum\limits_{n = 1}^{{n_{{\rm{occ}}}}} {\langle u_{k + \delta k}^{n'}|u_k^n\rangle {\Psi _{n,k}}}  = \lambda {\Psi _{n',k + \delta k}},
\end{equation}
which can be further written in a more compact form
\begin{equation}
{F_k}{\Psi _k} = \lambda {\Psi _{k + \delta k}}
\end{equation}
where $\Psi _k=(\Psi _{1,k}, \Psi _{2,k},...,\Psi _{n_{\rm{occ}},k})^{\rm{T}}$, $[F_k]_{n',n} =\langle u_{k + \delta k}^{n'}|u_k^n\rangle $.
Repeating the iterative relation for $L$ times, we have
\begin{equation}
{{\cal W}_{k + 2\pi  \leftarrow k}} {\Psi _k}{\rm{ = }}{\lambda ^L}{\Psi _k},
\end{equation}
which reveals that the Wilson loop is related to the eigenvalues of the projected position operator.
We may obtain the bulk polarization directly through the projected position operator.
In fact, the eigenstates of projected position operator are Wannier states, while its eigenvalues are center of mass of Wannier states \citep{PhysRevB.26.4269,niu1991theory,RevModPhys.84.1419}.
To summarize, one may obtain the polarization from projected position operators directly in real space.
This is beneficial for investigating the disorder system since the translation symmetry is broken and the Wilson loop method Eq.~\eqref{eqn:Wilson_loop} is not applicable.

\section{Winding number of 1D chiral-symmetric topological insulator}
\label{sec:Winding_Number}
\subsection{Chiral-symmetric system and sigular-value decomposition}
\label{sec:chiral_symmetric_system}
In this section, we shall firstly review some properties of the chiral-symmetric system.
Due to the special form of the chiral-symmetric Hamiltonian, we will introduce the singular value decomposition (SVD) for the Hamiltonian and construct a real-space representation of the winding number.
A lattice with chiral symmetry can be classified into two kinds of sublattices, namely $A$ and $B$.
Thus, the chiral symmetry is also called the sublattice symmetry.
The Hilbert space of the system can be written as the direct sum of the two subspaces ${\cal H} = {{\cal H}_A} \oplus {{\cal H}_B}$.
The chiral symmetry manifests that $\Gamma  H{\Gamma} =  - H$, where
\begin{equation}
\Gamma  = \sum\limits_{l,\alpha  \in A} {|l,\alpha \rangle \langle l,\alpha |}  - \sum\limits_{l,\beta  \in B} {|l,\beta \rangle \langle l,\beta |}  .
\end{equation}
In the canonical representation, where the chiral operator $\Gamma$ is diagonal, the Hamiltonian has the following structure
\begin{equation}
\label{eqn:chiral_Ham}
H = \left( {\begin{array}{*{20}{c}}
0&h\\
{{h^\dag }}&0
\end{array}} \right)
\end{equation}
where $h$ is a $L_A \times L_B$ matrix.
Here, $L_A$ and $L_B$ are respectively the total numbers of $A$ and $B$ sublattices.
By decomposing the eigenstates into two sectors $|\psi_n\rangle = (\psi^A_n, \psi^B_n)^T$, the eigenvalue equation $H|\psi_n\rangle=E_n|\psi_n\rangle$ leads to the coupled equations
\begin{equation}
\label{eqn:h_coupled_eqn}
\begin{array}{l}
h{\psi ^B_n} = E{\psi ^A_n}\\
{h^\dag }{\psi^A_n} = E{\psi ^B_n} ,
\end{array}
\end{equation}
which can be further written as
\begin{equation}
\label{eqn:eigval_coupled_eqns}
\begin{array}{l}
\left( {h{h^\dag }} \right)\psi _n^A = {E^2}\psi _n^A\\
\left( {{h^\dag }h} \right)\psi _n^B = {E^2}\psi _n^B
\end{array}
\end{equation}
Now we may obtain $\psi_n^A$ and $\psi_n^B$ by calculating the eigenvectors of ${h{h^\dag }} $ and ${{h^\dag }h}$ respectively.
Note that both $ (\psi^A_n, \pm \psi^B_n)^T$ are the eigenvectors of Hamiltonian \eqref{eqn:chiral_Ham}, and they have opposite eigenenergies.
This is a consequence of the chiral symmetry.
The expression of Eq.~\eqref{eqn:eigval_coupled_eqns} reminds us of the singular value decomposition (SVD).
We can make the SVD for the off-diagonal block $h = U_A\Sigma {U_B^{ - 1}}$, in which $U_A$ and $U_B$ are both unitary matrices, and $\Sigma$ is a diagonal matrix.
The diagonal elements of $\Sigma$ are called singular values.
We note that
\begin{eqnarray}
U_A^{ - 1}h{h^\dag } {U_A} &=&{\Sigma ^2}, \nonumber \\
U_B^{ - 1}{h^\dag }h{U_B} &=& {\Sigma ^2},
\end{eqnarray}
which reveals that $U_A$ and $U_B$ respectively diagonalize $h{h^\dag } $ and ${h^\dag }h$.
Therefore, the singular values are identified as the non-negative eigenenergies of the system.
The column of unitary matrices $U_A$ and $U_B$ are respectively the eigenvectors $\psi_n^A$ and $\psi_n^B$ (up to a normalization factor $1/\sqrt{2}$).
We have not assumed the numbers of the two kinds of sublattices $L_A,L_B$ in the above derivation.
Actually, $L_A$ and $L_B$ can be different, and the matrix $h$ is not necessarily squared.
If $L_A\ne L_B$, there must be at least $|L_A-L_B|$ zero singular values.
However, throughout this paper, we only consider the numbers of $A$ and $B$ sublattices are equal.
More precisely, we are interested in the situation where $h$ is not singular.
This further requires the system is gapped at $E=0$.
When all singular values are non-zero, one may deform the singular values to arbitrary positive values and still maintain the same eigenstates.
Hence, it is convenient to set $\Sigma$ to be identity.
We denote the new Hamiltonian as
\begin{equation}
\label{eqn:flattened_Ham_real}
Q = \left( {\begin{array}{*{20}{c}}
0&q\\
{{q^{ - 1}}}&0
\end{array}} \right),
\end{equation}
where $q=U_A U_B^{-1}$ is a unitary matrix.
Now the energy spectrum of the Hamiltonian becomes completely flat and only takes values of $+1$ and $-1$.
$Q$ is called the flattened Hamiltonian \citep{RevModPhys.88.035005}.
Later we will see the SVD is convenient for calculating and understanding the winding number.

\subsection{Winding number in momentum space}
When translation symmetry is present, we can work in momentum space by introducing the Fourier transformation
\[|k,\alpha \rangle  = \int {\frac{{dk}}{{2\pi }}\exp \left( {ikl} \right)|l,\alpha \rangle }. \]
Then we can block diagonalize the flattened Hamiltonian \eqref{eqn:flattened_Ham_real} as
\begin{equation}
{Q(k)} = \left( {\begin{array}{*{20}{c}}
0&{q\left( k \right)}\\
{q{{\left( k \right)}^{-1} }}&0
\end{array}} \right) .
\end{equation}
The eigenstates can be written as $|{\psi _{n,k}}\rangle  = \sum_\alpha  {u_{n,k}^\alpha |k,\alpha \rangle } $.
There is a map from Brillouin Zone to the unitary matrices $q(k)$.
The map is classified by the first homotopy group ${\pi _1}[U(n)] \cong \mathbb{Z}$ and characterized by the winding number \citep{RevModPhys.88.035005,ryu2010topological}.
The winding number in 1D can be calculated via \citep{RevModPhys.88.035005,ryu2010topological}
\begin{equation}
\label{eqn:winding_number_k}
\nu = \frac{i}{{2\pi}}\int_{ - \pi }^\pi  {dk{{\rm{Tr}}\left[ {{q}\left( k \right)^{ - 1}{\partial _k}q\left( k \right)} \right]} } .
\end{equation}
By inserting $q(k)=U_A(k)U_B^{-1}(k)$ introduced in previous subsection into Eq.~\eqref{eqn:winding_number_k}, we find
\begin{eqnarray}
\label{eqn:winding_number_SVD_k}
\nu &=&\frac{i}{{2\pi }}\int_{ - \pi }^\pi  {dk{{\rm{Tr}}\left[ {{U_A}\left( k \right)^{ - 1}{\partial _k}U_A\left( k \right)} \right]} }  \nonumber \\
&& \;\; - \frac{i}{{2\pi }}\int_{ - \pi }^\pi  {dk{{\rm{Tr}}\left[ {U_B\left( k \right)^{ - 1}{\partial _k}{U_B}\left( k \right)} \right]} } ,
\end{eqnarray}
where we have used a fact that $\mathrm{Tr}\left({U_q}^{ - 1}{\partial _q}{U_q} \right)=  - \mathrm{Tr} \left({U_q}{\partial _q}U_q^{ - 1}\right)$ when $U_q$ is a unitary matrix.
As mentioned above, the column of unitary matrix  $U_\sigma(k)$ is the eigenvector $u_{n,k}^\sigma \; (\sigma  = A,B)$ up to a normalization factor.
One can find that Eq.~\eqref{eqn:winding_number_SVD_k} is exactly the ``skew" polarization.
The above expression can be considered as the difference of the polarization between two sublattices.
In other words, the winding number (divided by 2) measures the difference of polarization between $A$ and $B$ sublattices.
In addition, one can find that the summation of the winding of $U_A(k)$ and $U_B(k)$ divided by 2 mode 1 leads to the polarization.
This implies a relation between polarization and winding number in chiral-symmetric topological insulators: $p=\nu/2 \;\mathrm{mod}\;1$.
\subsection{Winding number in real space}
%
Recently, the winding number is generalized from momentum-space formula [Eq.~\eqref{eqn:winding_number_k}] to real-space formula~\citep{PhysRevLett.113.046802, meier2018observation, PhysRevB.89.224203}.
The idea introduced in Ref.~\citep{PhysRevLett.113.046802} is to replace the integral and the derivative versus quasi-momentum $k$ by its real-space representation.
For a 1D system, the real-space winding number reads \citep{PhysRevLett.113.046802}
\begin{equation}
\label{eqn:real_space_windingNumber_original}
\nu  =  {\cal T}\left\{ {{Q_{BA}}\left[ {X,{Q_{AB}}} \right]} \right\} ,
\end{equation}
where $\cal T$ refers to trace per volume, ${Q_{BA}} = {\Gamma _B}Q{\Gamma _A},{Q_{AB}} = {\Gamma _A}Q{\Gamma _B}$, and ${{\Gamma }_{\sigma }}=\sum\nolimits_{l,\alpha \in \sigma }{|l,\alpha \rangle \langle l,\alpha |} $ is the projector onto the $\sigma=A,B$ subspaces \citep{meier2018observation}.
The real-space formula is still valid even in the presence of disorder given that the chiral symmetry is preserved \citep{PhysRevLett.113.046802,PhysRevB.89.224203}.
Here, we present a quite different formula to calculate the winding number in real space.
Later, we will prove that these two formulas are equivalent in the thermodynamic limit at half filling.
As stated in the previous section, the winding number is related to the relative polarization of $A$ and $B$ sublattice.
We shall follow the projected position operator approach described in Sec.~\ref{sec:bulk_polarization_and_projected_position_operator} to derive the winding number in real space.
To illustrate our idea, we consider a finite system with $L$ cells and discretize the integral in Eq.~\eqref{eqn:winding_number_SVD_k} as
\begin{equation}
\label{eqn:discretized_skew_polarization}
\nu  = \frac{1}{{2\pi i}}\sum\limits_k {{\rm{Tr}}\left[ {{\rm{log}}\left( {{F}_k^A{{ {{F}_k^B} }^\dag }} \right)} \right]},
\end{equation}
where $F_k^\sigma  = U_\sigma ^\dag \left( {k } \right){U_\sigma }\left( k - \delta k \right),\;\sigma  = A,B$ and $k=2n\pi/L,n\in \mathbb{Z}$ (see Appendix~\ref{appendix:derivation_discrete_winding_number} for detailed derivation).
Since the $n$-th column of $U_\sigma(k)$ is the vector $|u^\sigma_{n,k}\rangle$, it can be found that the matrix elements of $F_k^\sigma$ are ${\left[ {F_k^\sigma } \right]_{m,n}} = \langle u_{m,k}^\sigma |u_{n,k - \delta k}^\sigma \rangle $.
Next, we propose the following equivalent formula to calculate the winding number
\begin{equation}
\label{eqn:real_space_windingNumber}
\nu  = \frac{1}{{2\pi i}}{\rm{Tr}}\left[ {\log \left( {{{ \mathcal{X}}_A} \mathcal{X}_B^{ - 1}} \right)} \right],
\end{equation}
where ${{ \mathcal{X}}_\sigma } = U_\sigma ^{ - 1} \Gamma_\sigma \mathcal{X} \Gamma_\sigma {U_\sigma }$ ($\sigma=A,B$) are unitary matrices.
${{ \mathcal{X}}_\sigma }$ can be considered as the position operator projected onto the $\sigma$ sector of the eigenstate in the occupied band.
The quantum-mechanical position operator can be chosen as
\begin{equation}
\mathcal{X}=\sum\limits_{l,\alpha \in A,\beta \in B}{{{e}^{i\frac{2\pi }{L}l}}\left( |l,\alpha \rangle \langle l,\alpha |+|l,\beta \rangle \langle l,\beta | \right)},
\end{equation}
Then this operator have the same form when it is projected to the $A$ and $B$ sectors.
For convenience, we denote the projected operators as $\mathcal{\tilde X}\equiv\Gamma_{A}\mathcal{X}{{\Gamma }_{A}}={{\Gamma }_{B}}\mathcal{X}{{\Gamma }_{B}}$ in the following context.
To prove the equivalence between Eqs.~\eqref{eqn:real_space_windingNumber} and \eqref{eqn:discretized_skew_polarization}, we note that
\begin{eqnarray}
\label{eqn:translation_op_mat_element}
\left[ {{ \mathcal{X}}_\sigma }\right]_{(m,k'),(n,k)} &=& \langle \psi _{m,k'}^\sigma| \mathcal{X}|\psi _{n,k}^\sigma\rangle  \nonumber \\
&=&{\delta _{k',k + \delta k}}\langle u_{m,k'}^\sigma|u_{n,k}^\sigma\rangle,
\end{eqnarray}
where $|\psi _{n,k}^\sigma \rangle  = {\sum _{\alpha  \in \sigma }}u_{n,k}^\sigma |k,\alpha \rangle $ is the $ \sigma$ sector of eigenstate (up to $1/\sqrt{2}$ normalization factor).
Then, we have
\begin{widetext}
\begin{eqnarray}
{\left[ {{{\mathcal{X}}_A} \mathcal{X}_B^{ - 1}} \right]_{(m,k'),(l,k'')}}&=&\sum\limits_{n,k} {{{\left[ {{{ \mathcal{X}}_A}} \right]}_{(m,k'),(n,k)}}{{\left[ { \mathcal{X}_B^{ - 1}} \right]}_{(n,k),(l,k'')}}}  \nonumber \\
&=&\sum\limits_{n,k} {{\delta _{k',k + \delta k}}{\delta _{k,k'' - \delta k}}\langle u_{m,k'}^A|u_{n,k}^A\rangle \langle u_{n,k}^B|u_{l,k''}^B\rangle }  \nonumber \\
&=&  {\delta _{k',k''}}\sum\limits_{n} {\langle u_{m,k'}^A|u_{n,k' - \delta k}^A\rangle \langle u_{n,k''-\delta k}^B|u_{l,k''}^B\rangle } \nonumber \\
&=& {\delta _{k',k''}} \sum\limits_{n}{{\left[ {F_{k'}^A} \right]_{m,n}}{\left[ {\left( {F_{k'}^B} \right)^\dag } \right]_{n,l}}}.
\end{eqnarray}
\end{widetext}
Now we can see that matrix ${{{ \mathcal{X}}_A} \mathcal{X}_B^{ - 1}}$ has a block-diagonal structure.
Each block is associated with certain quasi-momentum $k$, and is exactly equal to the matrix $ {{F}_k^A{{ {{F}_k^B} }^\dag }} $.
Therefore, one can immediately recognize that Eqs.~\eqref{eqn:real_space_windingNumber} and \eqref{eqn:discretized_skew_polarization} are equivalent.
Note that Eq.~\eqref{eqn:discretized_skew_polarization} should reproduce a strictly quantized winding number $\nu \in \mathbb{N}$.
This can be proved by noting that $\det \left( {{\mathcal{X}_A}\mathcal{X}_B^{ - 1}} \right)= 1$, since $U_\sigma$ and $\mathcal{\tilde X}$ are unitary matrices.
Then, tracing the logarithm in Eq.~\eqref{eqn:real_space_windingNumber} gives an integer multiple of $2\pi i$, and therefore the resulting winding number is an integer number.
We have to emphasize that this quantization occurs only when the system is half-filled, i.e. the Fermi energy lies in the bandgap.
Generally, when the Fermi surface lies in the bands (i.e. the filling is less than one half), the momentum-space winding number in Eq.~\eqref{eqn:winding_number_k} is ill-defined.
However, the real-space formula~\eqref{eqn:real_space_windingNumber} enables us to calculate the `winding number' at arbitrary fractional filling less than one half.
As mentioned above, each column of $U_\sigma$ is the $\sigma$ sector of the eigenstates.
Thus, one may select certain columns of $U_\sigma$ to construct a reduced matrix $\tilde{U}_\sigma$, and then use Eq.~\eqref{eqn:real_space_windingNumber} to calculate the `winding number'.
Physically, it can be understood as projecting the position operator onto a certain subspace.
For example, we may choose the eigenstates whose energies are below the Fermi energy $E_n < E_\mathrm{F}$ and calculate the corresponding fractional winding number via Eq.~\eqref{eqn:real_space_windingNumber}.
As $\tilde{U}_\sigma$ is no more a unitary matrix after the reduction, the result may give a fractional value.
We will verify this method numerically in Sec.~\ref{sec:WN_fractional_filling}.
In addition, we shall show that our real-space representation of the winding number can be written in a form of the Bott index.
By transforming ${{{\mathcal{X}}_A} \mathcal{X}_B^{ - 1}}$ via unitary matrix $U_B$ and using $q=U_A U_B^{-1}$, we have
\begin{eqnarray}
\label{eqn:qxqx}
{U_B}\left( {{\mathcal{X}_A}\mathcal{X}_B^{ - 1}} \right)U_B^{ - 1} &=& {U_B}\left( {U_A^{ - 1}\mathcal{\tilde X}{U_A}U_B^{ - 1}{\mathcal{X}^{ - 1}}{U_B}} \right)U_B^{ - 1} \nonumber \\
&=&\left( {U_B}U_A^{ - 1}\right)\mathcal{\tilde X}\left({U_A}{U_B^{-1}}\right){\mathcal{\tilde X}^{ - 1}}  \nonumber \\
&=& {q^{ - 1}}\mathcal{\tilde X}q{\mathcal{\tilde X}^{ - 1}} .
\end{eqnarray}
Hence, Eq.~\eqref{eqn:real_space_windingNumber} can be written as
\begin{eqnarray}
\label{eqn:qxqx1}
\nu  &=& \frac{1}{{2\pi i}}{\rm{Tr}}\log \left( {{q^{ - 1}}\mathcal{\tilde X}q{\mathcal{\tilde X}^{ - 1}}} \right) \nonumber \\
&=& \mathrm{Bott}(q^{-1},\mathcal{\tilde X}),
\end{eqnarray}
which is exactly the form of Bott index introduced in \citep{exel1991invariants,hastings2010almost}.
In previous works, the Bott index is related to the real-space Chern number of the 2D topological insulator.
To the best of our knowledge, the Bott index has not been applied to the real-space winding number of 1D chiral-symmetric topological insulators.
Although the Bott indices for the Chern number and winding number have similar forms, they are fundamentally distinct.

\subsection{Winding number defined through twisted boundary condition}
In the previous subsection, we obtain a real-space representation of the winding number for the 1D chiral-symmetric topological insulator.
However, some of the properties, such as the self-averaging nature and the bulk-edge correspondence, are still vague.
In this subsection, we introduce the winding number defined through the twisted boundary condition (TBC).
Then, we prove the bulk-edge correspondence for the TBC winding number, which indicates that the TBC winding number is self-averaging in the presence of disorder in the thermodynamic limit.
Next, we will prove that our real-space representation of winding number Eq.~\eqref{eqn:real_space_windingNumber} is equivalent to the TBC winding number.
The TBC manifests that the two ends of the 1D lattice are glued together, but the particle will gain a phase $\Phi$ when they move through the boundary.
Thus, the TBC is also called the generalized periodic boundary condition \citep{Niu_1984, PhysRevB.31.3372}.
%
%
The TBC can be equivalently expressed as a result of the magnetic flux $\Phi$ piercing through the periodic chain.
General 1D Hamiltonian under TBC can be written as
\begin{eqnarray}
\label{eqn:TBC_Hamiltonian_boundary}
 && H\left( \Phi  \right)=-\sum\limits_{\alpha ,\beta ,n\le m}{t_{m,n}^{\alpha ,\beta }{{e}^{i{{\Phi }_{m,n}}}}c_{\alpha ,m}^{\dagger }{{c}_{\beta ,n}}+h.c.} \nonumber \\
 && {{\Phi }_{m,n}}=\left\{ \begin{matrix}
   \Phi , & \langle m,n\rangle \ \text{cross the boundary};  \\
   0, & \text{otherwise}  .\\
\end{matrix} \right. ,
\end{eqnarray}
in which $c^\dag_{\alpha,m}\; (c_{\alpha,m})$ is the creation (annihilation) operator of the $m$-th cell, $\alpha$ is the index of sublattice, and $t_{m,n}^{\alpha ,\beta }$ is the corresponding tunneling strength.

\subsubsection{Winding number and bulk-edge correspondence}
The bulk-edge correspondence is a well-known principle in topological band theory.
Non-trivial bulk topological invariant will lead to gapless excitations in the ground state at the edges.
There have been some rigorous mathematical proofs of bulk-edge correspondence in 1D chiral-symmetric topological insulator \citep{prodan2016bulk,graf2018bulk,shapiro2020bulk}.
Here, we use the TBC and follow the idea in Ref.~\citep{PhysRevB.74.045125} to derive this principle.
When the system possesses the chiral symmetry, as mentioned before, the flattened Hamiltonian is parametrized by the flux $\Phi$ 
\begin{equation}
Q\left( \Phi  \right)=\left( \begin{matrix}
   0 & q\left( \Phi  \right)  \\
   {{q}^{-1}}\left( \Phi  \right) & 0  \\
\end{matrix} \right) .
\end{equation}
In Ref.~\citep{PhysRevB.98.155137}, it has been proved that the excitation gap will not be affected by the twist angle $\Phi$ in the thermodynamic limit.
This means $q\left( \Phi  \right)$ is non-singular for arbitrary $\Phi \in [0, 2\pi]$ as long as the chiral-symmetric system is gapped at $E=0$.
Then, the TBC winding number can be safely defined as
\begin{equation}
\label{eqn:TBC_winding_Phi}
\tilde \nu = \frac{1}{2\pi i}\int\limits_{0}^{2\pi }{d\Phi \ \text{Tr}\left[ {{q}^{-1}}\left( \Phi  \right){{\partial }_{\Phi }}q\left( \Phi  \right) \right]}.
\end{equation}
Next, we show that non-trivial winding number $\tilde \nu\ne 0$ under PBC results in the zero-energy modes, and the number of zero-energy modes is twice the winding number.
Inspired by Ref.~\citep{PhysRevB.74.045125}, we modify the boundary condition by adding a parameter $\eta \in [0,1]$ onto the tunneling which crosses the boundary
\begin{equation}
t_{m,n}^{\alpha ,\beta }\left( \eta  \right)=\left\{ \begin{matrix}
   \eta t_{m,n}^{\alpha ,\beta }, & \langle m,n\rangle \ \text{cross the boundary}  \\
   t_{m,n}^{\alpha ,\beta }, & \text{otherwise}  \\
\end{matrix} \right.  .
\end{equation}
The system has open boundary when $\eta = 0$, and restores the usual TBC when $\eta = 1$.
Now the Hamiltonian are parameterized by $(\Phi, \eta)$.
Then, we introduce the $U(1)$ phase field
\begin{equation}
z\left( \Phi ,\eta  \right)=\det q\left( \Phi ,\eta  \right) \in U(1).
\end{equation}
Provided $z\left( \Phi ,\eta  \right)\ne  0$, the TBC winding number Eq.~\eqref{eqn:TBC_winding_Phi} can be written as
\begin{equation}
\label{eqn:TBC_winding_phase_field}
\tilde{\nu }=\frac{1}{2\pi i}\oint\limits_{\eta =1}{{{z}^{-1}}dz}.
\end{equation}
For non-trivial winding number $\tilde{\nu}\ne 0$, Eq.~\eqref{eqn:TBC_winding_phase_field} implies some poles of $z$ reside in the circle of $\eta = 1$, and the number of the poles should be equal to the absolute value of winding number.
Recall that the twist angle will not affect the excitation gap in the thermodynamic limit, the system should be either gapped or gapless at $E=0$ for arbitrary twist angle $\Phi \in [0, 2\pi]$.
This argument also holds for $\eta < 1$.
Hence, there should be an infinite number of poles inside the circle if $z \left(\Phi, \eta  \right) = 0$ for $0 < \eta < 1$, which is impossible when $\nu$ is well-defined and the system is away from the phase transition.
Consequently,  the poles of $z \left( \Phi ,\eta  \right)$ can only occur at $\eta = 0$, which corresponds to the open boundary condition.
This proof is similar to the discussion about the bulk-edge correspondence of the quantum Hall effect in Ref.~\citep{PhysRevB.74.045125}.
Then, we prove that the appearance of the zero-energy modes is associated to the non-trivial winding number.
Note that $z\left( \Phi ,\eta  \right)=\det q\left( \Phi ,\eta  \right)$ is proportional to the following products
\begin{equation}
z\left( \Phi ,\eta  \right)=\det q\left( \Phi ,\eta  \right) \propto  \prod\limits_{n}{{{\xi }_{n}\left( \Phi ,\eta  \right)}}
\end{equation}
in which $\{\xi_n \left( \Phi ,\eta  \right)\}$ are the singular values of $q\left( \Phi ,\eta  \right)$.
We can learn from the above discussions that the number of poles is exactly equal to the number of zero singular values.
Meanwhile, as mentioned in Sec. \ref{sec:chiral_symmetric_system}, the singular values of the off-diagonal block $h$ are half of the eigenvalues of the corresponding chiral-symmetric Hamiltonian.
Therefore, we can conclude that the number of zero-energy modes is twice the winding number under open boundary condition.
Since we have assumed the system is gapped under PBC, the zero-energy modes are the in-gap modes and should be localized at the edge, which is known as the bulk-edge correspondence.
The above considerations are still valid if we replace the flattened Hamiltonian $Q$ by the original Hamiltonian $H$.
Importantly, this bulk-edge correspondence holds for the disordered case given that chiral symmetry persists.
The bulk-edge correspondence also implies that the TBC winding number is self-averaging in the thermodynamic limit and away from the phase transition point.

\subsubsection{Equivalence of the TBC winding number and the real-space representation of the winding number}
Now, we would like to prove that our real-space representation of winding number Eq.~\eqref{eqn:real_space_windingNumber} is equivalent to the TBC winding number in the thermodynamic limit at half filling.
The TBC Hamiltonian \eqref{eqn:TBC_Hamiltonian_boundary} can be transformed to \citep{PhysRevB.98.155137, PhysRevX.8.021065}
\begin{eqnarray}
\label{eqn:TBC_Ham_periodic}
\tilde{H}\left( \Phi  \right)&=&{{\mathcal{U}}_{\Phi }}H\left( \Phi  \right){{\mathcal{U}}_{\Phi }}^{-1} \nonumber \\
&=&-\sum\limits_{\alpha ,\beta ,n\le m}{t_{m,n}^{\alpha ,\beta }{{e}^{i\frac{\Phi }{L}\left( m-n \right)}}c_{\alpha ,m}^{\dagger }{{c}_{\beta ,n}}+h.c.}
\end{eqnarray}
where 
\begin{equation}
{{\mathcal{U}}_{\Phi }}={{e}^{i\frac{\Phi }{L}\hat{X}}},\ \ \hat{X}=\sum\limits_{\alpha ,m}{mc_{\alpha ,m}^{\dagger }{{c}_{\alpha ,m}}}.
\end{equation}
We shall use the tilde notation to distinguish these two unitarily equivalent Hamiltonians in the following discussions.
For a sufficiently large system $L\to \infty$, and assuming the range of tunneling is finite, we can expand Eq.~\eqref{eqn:TBC_Ham_periodic} up to the leading order of $ \Phi/L$
\begin{eqnarray}
\label{eqn:H_Phi_expansion}
&&\tilde{H}\left( \Phi  \right)=H\left( 0 \right)+\frac{\Phi }{L}\mathcal{J}+O\left( \frac{1}{{{L}^{2}}} \right)	\nonumber \\
&&\mathcal{J}=i\sum\limits_{\alpha ,\beta ,n\le m}{\left[ \left( m-n \right)t_{m,n}^{\alpha ,\beta }c_{\alpha ,m}^{\dagger }{{c}_{\beta ,n}}-h.c. \right]}.
\end{eqnarray}
One may notice that $\mathcal{J}$ is the current operator.
Similarly, we can expand $\tilde q(\Phi)$ up to the leading order of $\epsilon = \Phi/L$
 \begin{equation}
 \label{eqn:q_epsilon}
 \tilde q\left( \epsilon  \right)=q\left( 0 \right)+\epsilon {{\left[ {{\partial }_{\epsilon }} \tilde q\left( \epsilon  \right) \right]}_{\epsilon =0}}+O\left( {{\epsilon }^{2}} \right)
 \end{equation}
 where we have expressed $\tilde q(\Phi)$ as $\tilde q(\epsilon)$ for clarity.
 With this approximation, the winding number Eq.~\eqref{eqn:TBC_winding_Phi} can be written as
\begin{eqnarray}
\label{eqn:winding_number_epsilon}
 \nu &=&\frac{1}{2\pi i}\int\limits_{0}^{2\pi /L}{d\epsilon \ \text{Tr}\left[ {{\tilde q}^{-1}}\left( \epsilon  \right){{\partial }_{\epsilon }}\tilde q\left( \epsilon  \right) \right]} \nonumber \\
&=& \frac{1}{2\pi i}\int\limits_{0}^{2\pi /L}{d\epsilon \ \text{Tr}\left[ {{q}^{-1}}\left( 0 \right){{\left[ {{\partial }_{\epsilon }}\tilde q\left( \epsilon  \right) \right]}_{\epsilon =0}} \right]} +O\left( {{\epsilon }^{2}} \right) \nonumber \\
&=& \frac{1}{iL}\text{Tr}\left[ {{q}^{-1}}\left( 0 \right){{\left[ {{\partial }_{\epsilon }}\tilde q\left( \epsilon  \right) \right]}_{\epsilon =0}} \right]
\end{eqnarray}
where we have used the fact that $\left[ {{q}^{-1}}\left( 0 \right){{\left[ {{\partial }_{\epsilon }}\tilde q\left( \epsilon  \right) \right]}_{\epsilon =0}} \right]$ is independent on $\epsilon$.
On the other hand, noting that $\mathcal{U}_{2\pi} = \mathcal X$, there is $H\left( 0 \right)= H\left( 2 \pi \right)= \mathcal{X}^{-1} \tilde H \left( 2\pi  \right){{\mathcal{X}}}$ according to Eq.~\eqref{eqn:TBC_Ham_periodic}.
This relation can be also generalized to the flattened Hamiltonian $Q\left( 0 \right)=\mathcal{X}^{-1} \tilde Q \left( 2\pi  \right){{\mathcal{X}}}$. 
Since the matrix $q(\Phi)$ can be obtained from projecting the Q matrix via $\tilde q\left( \Phi  \right)={{\Gamma }_{A}}\tilde Q\left( \Phi  \right){{\Gamma }_{B}}$.
We can derive a similar relation for the $q(0)$ and $\tilde q(2\pi)$ 
\begin{eqnarray}
\label{eqn:q_0_q2pi_relation}
q\left( 0 \right) &=& {{\Gamma }_{A}} Q \left( 0 \right){{\Gamma }_{B}}={{\Gamma }_{A}}\mathcal{X}^{-1}\tilde Q\left( 2\pi  \right){{\mathcal{X}}}{{\Gamma }_{B}} \nonumber \\
&=& \left( {{\Gamma }_{A}}{{\mathcal{X}}^{-1}}{{\Gamma }_{A}} \right)\left[ {{\Gamma }_{A}}\tilde Q\left( 2\pi  \right){{\Gamma }_{B}} \right]\left( {{\Gamma }_{B}}\mathcal{X}{{\Gamma }_{B}} \right) \nonumber \\
&=& {{{\tilde{\mathcal{X}}}}^{-1}}\tilde q\left( 2\pi  \right)\tilde{\mathcal{X}}
\end{eqnarray}
where we have used the fact that $\mathcal{X}$ is diagonal in the position space and ${{\Gamma }_{\sigma }}\mathcal{X}={{\Gamma }_{\sigma }}\mathcal{X}{{\Gamma }_{\sigma }},\ \ {{\Gamma }^{2}_{\sigma }}={{\Gamma }_{\sigma }}$, ($\sigma=A,B$).
Let $\epsilon = 2\pi/L$ in Eq.~\eqref{eqn:q_epsilon}, we can combine it with Eq.~\eqref{eqn:q_0_q2pi_relation} and obtain the following relation
\begin{eqnarray}
 \label{eqn:XqX_q_2piL}
{\mathcal{\tilde X} }q\left( 0 \right)\mathcal{\tilde X}^{-1} &=&   \tilde q\left( \epsilon = \frac{2\pi}{L}  \right)  \nonumber \\
&=& q\left( 0 \right) + \frac{2\pi}{L} {{\left[ {{\partial }_{\epsilon }}\tilde q\left( \epsilon  \right) \right]}_{\epsilon =0}} + O(\frac{1}{L^2}).
\end{eqnarray}
Therefore, we can rewrite Eq.~\eqref{eqn:winding_number_epsilon} as
\begin{eqnarray}
\label{eqn:winding_number_trXqX}
\nu &=& \frac{1}{iL}\text{Tr}\left\{ {{q}^{-1}}\left( 0 \right){{\left[ {{\partial }_{\epsilon }}\tilde q\left( \epsilon  \right) \right]}_{\epsilon =0}} \right\} \nonumber \\
&=& \frac{1}{2\pi i}\text{Tr}\left\{ {{q}^{-1}}\left( 0 \right)\left[ {{\mathcal{\tilde X}}}q\left( 0 \right)\mathcal{\tilde X}^{-1}-q\left( 0 \right) \right] \right\} \nonumber \\
&=& \frac{1}{2\pi i}\text{Tr}\left( {{q}^{-1}}\mathcal{\tilde X}q{{\mathcal{\tilde X}}^{-1}}-I \right).
\end{eqnarray}
where $I$ is the identity matrix and we have dropped the dependence on the twist angle in the last row.
From Eq.~\eqref{eqn:XqX_q_2piL}, one can find that ${{q}^{-1}}\mathcal{\tilde X}q{{\mathcal{\tilde X}}^{-1}}$ is close to the identity for sufficiently large $L$.
In the thermodynamic limit $L\to \infty$, Eq.~\eqref{eqn:winding_number_trXqX} can be written as the matrix logarithm \citep{higham2008functions}
\begin{equation}
\nu =\frac{1}{2\pi i}\text{Tr}\log \left( {{q}^{-1}}\mathcal{\tilde X}q{{\mathcal{\tilde X}}^{-1}} \right)
\end{equation}
which is in agreement with Eq.~\eqref{eqn:qxqx1}.
Thus, we have proved that the real-space representation of the winding number is equivalent to the winding number defined through the TBC in the thermodynamic limit at half filling.
Generally speaking, to obtain the winding number defined through the TBC, one needs to change the twist angle $\Phi$ from $0$ to $2\pi$, which requires great computational resources.
Here, we have shown that the winding angle of $\tilde q(\Phi)$ changes linearly with the twist angle: $\arg \left[ \det  \tilde q\left( \Phi  \right) \right]\sim \Phi $.
It is such a kind of linear dependence that leads to our efficient real-space representation of the winding number.
Moreover, as discussed above, the TBC winding number should be a self-averaging quantity, and we can therefore conclude that our real-space representation of the winding number is also self-averaging.
In addition, we can further prove that the TBC winding number [Eq.~\eqref{eqn:TBC_winding_Phi}], as well as the real-space representation of the winding number [Eq.~\eqref{eqn:real_space_windingNumber}] in our work, are equivalent to the formula [Eq.~\eqref{eqn:real_space_windingNumber_original}] obtained in Ref.~\citep{PhysRevLett.113.046802}.
To show this, we note that in the thermodynamic limit, there is
\begin{eqnarray}
\left[ X,H\left( 0 \right) \right]&=&\sum\limits_{\alpha ,\beta ,n\le m}{\left[ \left( m-n \right)t_{m,n}^{\alpha ,\beta }c_{\alpha ,m}^{\dagger }{{c}_{\beta ,n}}-h.c. \right]} \nonumber \\
&=& - i \mathcal{J} ,
\end{eqnarray}
where $X=\sum\nolimits_{\alpha ,m}{mc_{\alpha ,m}^{\dagger }{{c}_{\alpha ,m}}}$.
On the other hand, the off-diagonal block $h(\Phi)$ matrix and its inverse read as
\begin{equation}
\tilde  h\left( \Phi  \right)={{\Gamma }_{A}}\tilde H\left( \Phi  \right){{\Gamma }_{B}},\;\; \tilde h\left( \Phi  \right)^{-1}={{\Gamma }_{B}}\tilde H\left( \Phi  \right)^{-1}{{\Gamma }_{A}} .
\end{equation}
Hence, using Eq.~\eqref{eqn:H_Phi_expansion}, the derivative with respect to $\Phi$ reads as
\begin{eqnarray}
{{\left[ {{\partial }_{\epsilon }}\tilde h\left( \epsilon  \right) \right]}_{\epsilon =0}}&=&L{{\left[ {{\partial }_{\Phi }}\tilde h\left( \Phi  \right) \right]}_{\Phi =0}} \nonumber \\
&=&{{\Gamma }_{A}}\mathcal{J} {{\Gamma }_{B}}\nonumber \\
&=& i{{\Gamma }_{A}} [X,H(0)] {{\Gamma }_{B}}.
\end{eqnarray}
The winding number [Eq.~\eqref{eqn:winding_number_epsilon}] can be equivalently written as 
\begin{equation}
\tilde{\nu}=\frac{1}{iL}h{{\left( 0 \right)}^{-1}}{{\left[ {{\partial }_{\epsilon }}\tilde h\left( \epsilon  \right) \right]}_{\epsilon =0}}=\frac{1}{L}{{\Gamma }_{B}}H^{-1}{{\Gamma }_{A}}\left[ X,H \right]{{\Gamma }_{B}},
\end{equation}
where we have dropped the dependence on the twist angle $\Phi$ in the last term for simplicity.
Then, it is tempting to replace the Hamiltonian by the flattened Hamiltonian, and we have
\begin{eqnarray}
\label{eqn:winding_QAB_equivalent}
\tilde{\nu} &=& \frac{1}{L}{{\Gamma }_{B}}Q{{\Gamma }_{A}}\left[ X,Q \right]{{\Gamma }_{B}}\nonumber \\
&=& \frac{1}{L}{{Q}_{BA}}\left[ X,{{Q}_{AB}} \right], 
\end{eqnarray}
in which we have used the relation $\Gamma_\sigma^2=\Gamma_\sigma$, $\Gamma_\sigma X=\Gamma_\sigma X \Gamma_\sigma$ ($\sigma=A,B$) and $Q^{-1}=Q$.
We find that Eq.~\eqref{eqn:winding_QAB_equivalent} is identical to Eq.~\eqref{eqn:real_space_windingNumber_original}.
Therefore, we can conclude that the TBC winding number [Eq.~\eqref{eqn:TBC_winding_Phi}], the real-space representation of the winding number [Eq.~\eqref{eqn:real_space_windingNumber}] in our work, and the real-space formula [Eq.~\eqref{eqn:real_space_windingNumber_original}] obtained in Ref.~\citep{PhysRevLett.113.046802}, are equivalent in the thermodynamic limit.
Note that Eq.~\eqref{eqn:winding_QAB_equivalent} can be also derived from Eq.~\eqref{eqn:winding_number_trXqX} by using the Baker-Campbell-Hausdorff formula to expand $\mathcal{\tilde X}q\mathcal{\tilde X}^{-1}$ up to the first term, but the convergence of the expansion seems to be unclear in that way. 

\subsection{Winding number defined through path connection}
Previously, we obtain the real-space winding number by considering the sublattice polarization.
We find that our formula can be expressed as the Bott index.
For the Bott index, there is an alternative method to express the winding number by a path connection \citep{exel1991invariants}, which clearly shows the meaning of winding in real space.
We shall give a simple illustration here.
Since unitary matrix is always diagonalizable, we can make an eigenvalue decomposition ${\mathcal{X}_A}\mathcal{X}_B^{ - 1}=T \mathcal X_{\mathrm{diag}}T^{-1}$ where $T$ is an unitary matrix and $\mathcal X_{\mathrm{diag}} = \mathrm{diag}\left\{ {{e^{i{\theta _1}}},{e^{i{\theta _2}}}, \ldots ,{e^{i{\theta _n}}}} \right\}$ is a diagonal matrix with all elements lying on an unit circle in complex plane $\mathbb{C}$.
Then, Eq.~\eqref{eqn:real_space_windingNumber} can be written as
\begin{equation}
\nu  =  \frac{1}{{2\pi i}}{\rm{Tr}}\left[ {\log \left( {{\mathcal{X}_{{\rm{diag}}}}} \right)} \right] = \frac{1}{{2\pi }}\sum\limits_k {{\theta _k}} .
\end{equation}
In other words, the information of the winding number is encoded in these phases.
Recall that the matrix logarithm is a multi-valued matrix function.
To `unwind' the logarithm in Eq.~\eqref{eqn:real_space_windingNumber}, we consider a continuous path (homotopy) $\phi :[0,1] \to {\mathrm{GL}}\left( {n, \mathbb{C}} \right)$ such that $\phi(0) = {\mathcal X_A}\mathcal X_B^{ - 1}$ and $\phi(1) = I$.
The function $\phi(r)$ can be written as
\begin{eqnarray}
\phi \left( r \right) &=& T{\mathcal{X}_{{\rm{diag}}}}\left( r \right){T^{ - 1}}  \nonumber \\
&=& T{\rm{diag}}\left\{ {{e^{i{\tilde \theta _1}\left( r \right)}},{e^{i{\tilde \theta _2}\left( r \right)}}, \ldots ,{e^{i{\tilde \theta _n}\left( r \right)}}} \right\}{T^{ - 1}}
\end{eqnarray}
where $\tilde \theta_k(r) \in [0,2\pi), r\in [0,1]$ is a continuous real function with $\tilde \theta_k(0)=\theta_k$ and $\tilde \theta_k(1)=0$.
The winding number can be defined as
\begin{equation}
\label{eqn:Generalized_winding_number}
\nu = -\frac{1}{2\pi}\int_0^1  {dr}\frac{\partial }{{\partial r}}\arg \left[ {\det \phi \left( r \right)} \right] ,
\end{equation}
which is indeed related to the real-space winding number since
\begin{eqnarray}
 &-&\frac{1}{2\pi}  \int_0^1  {dr}\frac{\partial }{{\partial r}}\arg \left[ {\det \phi \left( r \right)} \right]  \nonumber \\
&&   \;\;\;\;  =-\frac{1}{2\pi}\sum_k {\int_0^1 {dr\frac{\partial }{{\partial r}}{{\tilde \theta }_k}\left( r \right)} }  = \frac{1}{2\pi}  \sum\limits_k {{\theta _k}} ,
\end{eqnarray}
where we have used $\int_0^1 {dr\frac{\partial}{{\partial r}}{\tilde \theta _k}\left( r \right)}  = -\theta_k$.
When the winding number is non-trivial $\nu>0$, the continuous path $\phi$ connects different branches of complex logarithm.
The winding number can only be changed discontinuously when the system undergoes a phase transition.
This is accompanied by the sudden changes of some phase $\theta_k$, and the function $\mathrm{arg} [\det \phi(r)]$ becomes discontinuous and indifferentiable.
Note that $\det \phi(r)$ may be still continuous and well-defined.
The singularity of $\mathrm{arg} [\det \phi(r)]$ is due to the multi-valued nature of the argument.
For example, when a phase grows continuously from $\theta_0$ to $\theta_0 + 2\pi$, the complex function $e^{i\theta_0}$ is continuous, while its principal argument $\mathrm{arg} (e^{i\theta_0})$ is discontinuous.
We will show this later with numerical calculation.
In the presence of translation symmetry, we have proved that our formula~\eqref{eqn:real_space_windingNumber} is exactly equal to the momentum-space formula~\eqref{eqn:winding_number_k}.
Thus, it is natural that the winding number [Eq.~\eqref{eqn:Generalized_winding_number}] is also equivalent to the momentum-space formula [Eq.~\eqref{eqn:winding_number_k}].
Also, it is always possible to elaborate a path that the behavior of the winding number defined in Eq.~\eqref{eqn:Generalized_winding_number} is identical to the momentum-space winding number [Eq.~\eqref{eqn:winding_number_k}] when translation symmetry is present.
In other words, the loop $\det \phi(r)$ is homotopic to $\det h(k)$ and $\det q(k)$ in the presence of translation symmetry.

\section{Application to 1D BDI class model}
\label{sec:application_BDI_model}
In the previous section, we have introduced a representation of the real-space winding number.
Our formula ensures that the winding number is an integer at half-filling case.
For arbitrary fractional filling less than one-half, our formula is still applicable but does not necessarily give an integer number.
On the other hand, we define a winding number through a continuous path from the special unitary matrix to the identity matrix.
In this section, we will apply these arguments to a toy model in the 1D BDI class.
We shall also consider a disorder on tunneling to see the robustness and validity of the real-space winding number.
These calculations can be easily extended to the AIII class.

\subsection{Extended Su-Schrieffer-Heeger model and disordered tunneling}
\begin{figure}[h]
  \includegraphics[width = \columnwidth ]{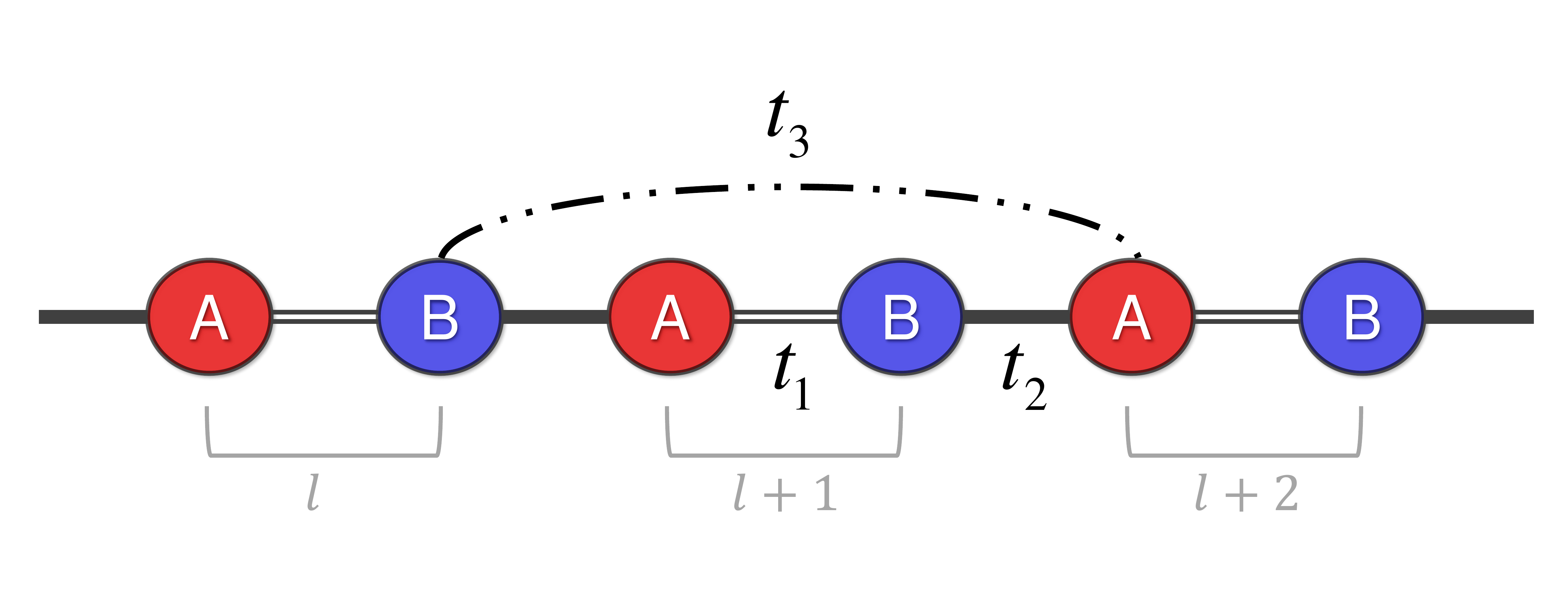}
  \caption{\label{fig:FIG_ExtSSH_Scheme}
 Schematic illustration of the extended SSH model.
 Besides the general nearest-neighbor tunneling in the SSH model, we include special long-range tunneling indicated by the dashed line.
  }
\end{figure}

\begin{figure}[h]
  \includegraphics[width = \columnwidth ]{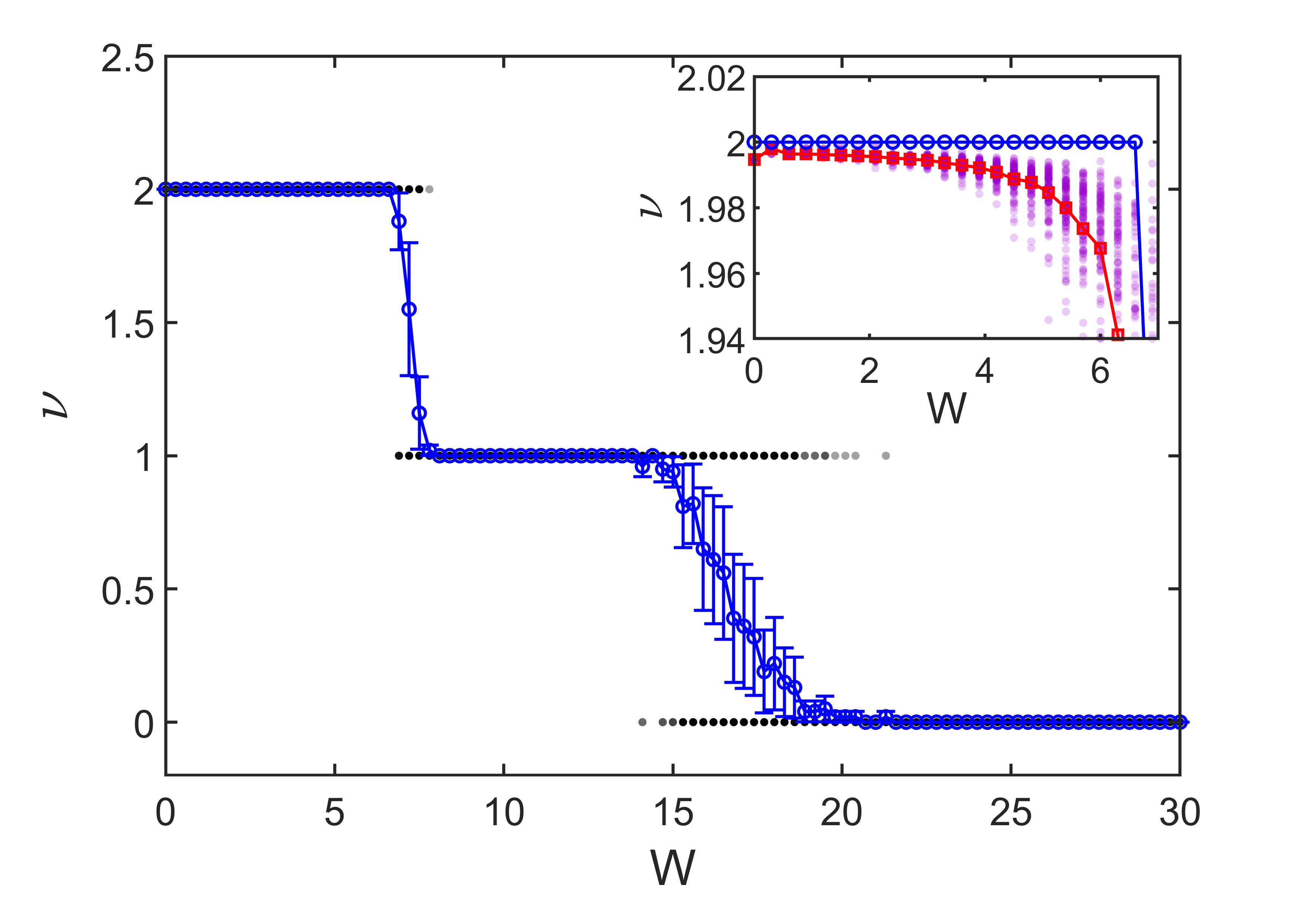}
  \caption{\label{fig:FIG_SSH_ScanW}
 Winding number $\nu$ as a function of disorder strength $W$ in disordered extended SSH model in Eq.~\eqref{eqn:extended_SSH_Ham}.
 Grey dots are the results from 100 random realizations based on Eq.~\eqref{eqn:real_space_windingNumber}.
 Blue circles are averaged over the grey dots.
 \emph{Inset}: Comparison between Eq.~\eqref{eqn:real_space_windingNumber_original} and Eq.~\eqref{eqn:real_space_windingNumber}.
 Light-purple dots are the results from 100 random realization based on Eq.~\eqref{eqn:real_space_windingNumber_original}.
 Red squares are the averaged results of the purple dots.
 The above calculations are implemented with $L=1001$ cells.
 The parameters are $t_1=0, t_2=1, t_3=-2$, and the configuration of disorder is $W=W_1=2W_2$, $W_3=0$.
  }
\end{figure}

\begin{figure}[h]
  \includegraphics[width = \columnwidth ]{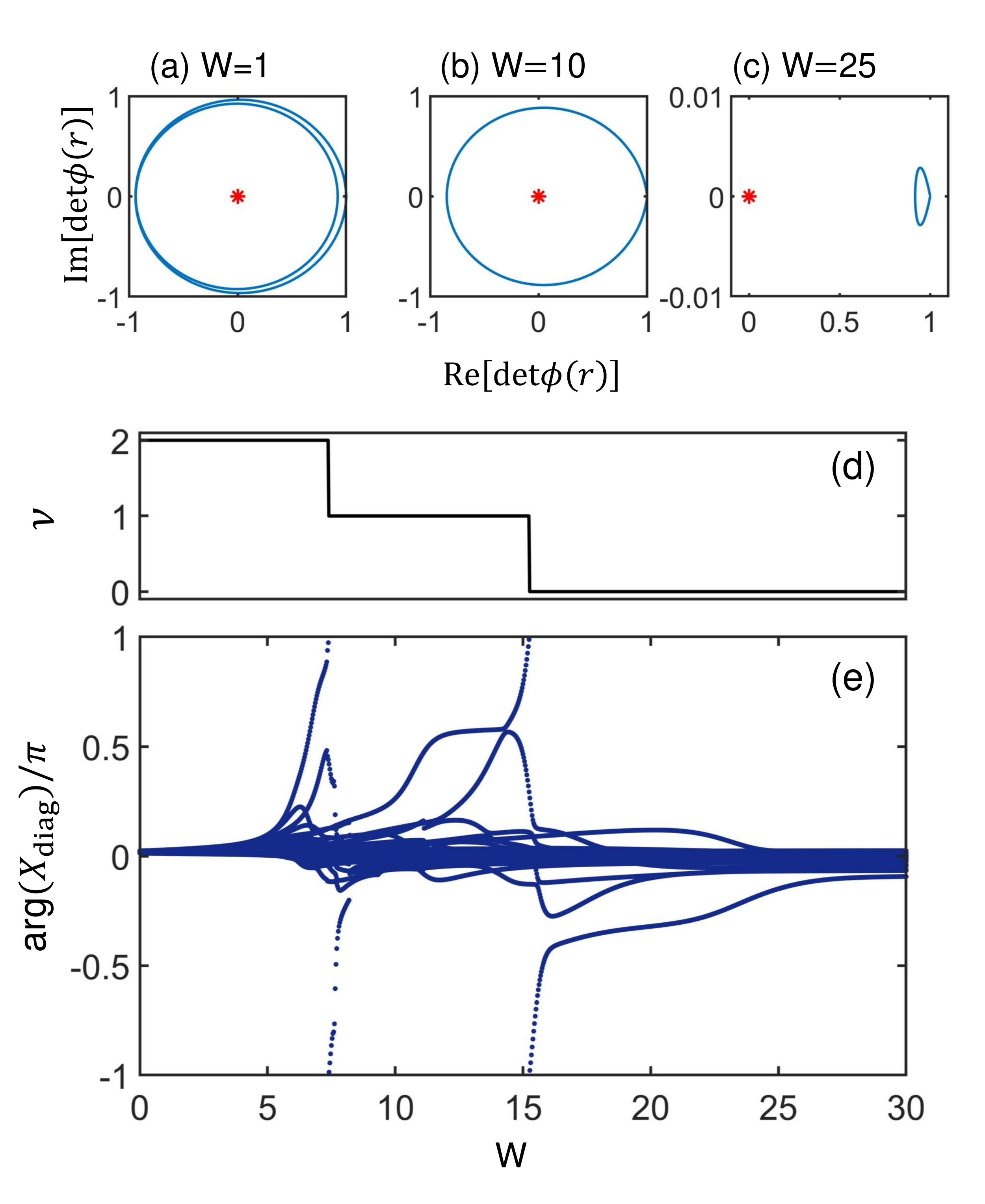}
  \caption{\label{fig:FIG_SSH_ScanW_Phase}
  (a-c) show the trajectory of $\det \phi(r)$ with $W=1, 10, 25$ in a single random realization.
 The red star represents the zero point.
 (d) shows the real-space winding number $\nu$ as a function of disorder strength $W$ given a certain set of random factor $\left\{ {{\varepsilon _{n,l}}} \right\}$ in a single random realization.
 (e) shows the argument $\left\{ {{\theta _k}} \right\} $ of the diagonal entries of $\mathcal{X}_\mathrm{diag}$ as a function of $W$ in the same random realization as (d).
 Other parameters are identical to the parameters in Fig.~\ref{fig:FIG_SSH_ScanW}.
  }
\end{figure}

We introduce a 1D lattice model described by the following second quantized Hamiltonian
\begin{equation}
\label{eqn:extended_SSH_Ham}
H =  - \sum\limits_l {\left( {{t_{1,l}}\hat a_l^\dag {\hat b_l} + {t_{2,l}}\hat a_{l + 1}^\dag {\hat b_l} + {t_{3,l}}\hat a_{l + 2}^\dag {\hat b_l}} \right)}  + H.c. ,
\end{equation}
where $\hat{a_l}^{\dag}$ ($\hat{b_l}^{\dag}$) creates a particle at the $\textrm{A}$ ($\textrm{B}$) sublattice of the $l$th cell.
 $t_{n,l}=t_n+W_n\varepsilon_{n,l},n=1,2,3$ is the tunneling strength, $\varepsilon_{n,l} \in (-1/2,1/2)$ is a random strength distributed uniformly, and $W_n$ is the strength of disorder.
This is an extended Su-Schrieffer-Heeger (SSH) model up to a next-next-nearest-neighbor (NNNN) term.
A schematic illustration of this toy model is shown in Fig.~\ref{fig:FIG_ExtSSH_Scheme}.
With this special NNNN tunneling, the system still preserves chiral symmetry.
According to the ten-fold classification, this model belongs to BDI class.
There exist topological phases characterized by winding number $\nu=0,1,2$ in this model.
As a benchmark, we adapt a similar configuration in Ref.~\citep{PhysRevB.89.224203}, and set $W=W_1=2W_2$, $W_3=0$ and $t_1=0, t_2=1, t_3=-2$.
We use the two approaches Eq.~\eqref{eqn:real_space_windingNumber_original} and Eq.~\eqref{eqn:real_space_windingNumber} to numerically calculate the winding number as a function of disorder strength $W$.
Here the position operator reads as $\mathcal{\hat X} = \exp[i\delta k \sum_i {l(\hat a^\dag_l \hat a_l + \hat b^\dag_l \hat b_l)}]$.
The results are shown in Fig.~\ref{fig:FIG_SSH_ScanW}.
It can be seen that the winding numbers obtained from Eq.~\eqref{eqn:real_space_windingNumber} stay quantized in each random realization (grey dots).
We have examined that the averaged results (blue circles) agree well with the averaged results obtained from Eq.~\eqref{eqn:real_space_windingNumber_original}.
Away from the phase transition point, there is almost no fluctuation of winding number because each realization is perfectly the same quantization. Around the phase transition point, we find that the fluctuation of winding number becomes large, which can be served as a signature of the topological phase transition in disorder systems.  
The phase transition point can be determined with higher accuracy as the lattice length increases. 
The scaling with different lattice lengths is presented in Appendix.~\ref{appendix:scaling}.
However, we note that the real-space winding number obtained from Eq.~\eqref{eqn:real_space_windingNumber_original} is not strictly quantized for the finite system.
As shown in the inset of Fig.~\ref{fig:FIG_SSH_ScanW}, the real-space winding number calculated via Eq.~\eqref{eqn:real_space_windingNumber_original} slightly deviates from the integer.
Next, based on Eq.~\eqref{eqn:Generalized_winding_number}, we use Eq.~\eqref{eqn:Generalized_winding_number} to graphically show the winding number in the presence of disorder.
Practically, we can simply assume
\begin{equation}
\phi \left( r \right) = {\left( {1 - r} \right){\mathcal{X}_A}\mathcal{X}_B^{ - 1} + r I}, \quad r\in [0,1].
\end{equation}
The determinant is
\begin{eqnarray}
\det \phi \left( r \right) &=& \det \left[ {\left( {1 - r} \right){\mathcal{X}_A}\mathcal{X}_B^{ - 1} + rI} \right] \nonumber \\
&=& \det \left\{ {T\left[ {\left( {1 - r} \right){\mathcal{X}_{{\rm{diag}}}} + rI} \right]{T^{ - 1}}} \right\} \nonumber \\
&=& \det \left[ {\left( {1 - r} \right){\mathcal{X}_{{\rm{diag}}}} + rI} \right] \nonumber \\
&=& \prod\limits_k {\left[ {\left( {1 - r} \right){e^{i{\theta _k}}} + r} \right]}  .
\end{eqnarray}
The graphical illustrations of real-space winding numbers are shown in Figs.~\ref{fig:FIG_SSH_ScanW_Phase} (a-c).
It can be seen that the winding around the singularity point ($\det \phi(r) = 0$) in the complex plane coincides well with the real-space winding number.
Besides, it can be found that $\det \phi \left( r \right) $ only takes zero value at $r=1/2$ and $\theta_k=\pi$.
As stated in the previous section, we may identify that $\theta_k=\pi$ is related to the discontinuous change of argument, and thus corresponds to phase transition.
%
To show this, we numerically calculate the arguments $\left\{ {{\theta _k}} \right\} $ from the diagonal entries of $X_\mathrm{diag}$ as a function of disorder strength $W$ in a \emph{single} random realization $\left\{ {{\varepsilon _{n,l}}} \right\}$.
As presented in Figs.~\ref{fig:FIG_SSH_ScanW_Phase} (d-e), given a random realization $\left\{ {{\varepsilon _{n,l}}} \right\}$, all the phases $\left\{ {{\theta _k}} \right\} $ change continuously with disorder strength except for the phase transition point.
This is in agreement with our discussion.

\subsection{Winding number under different filling factors}
\label{sec:WN_fractional_filling}
\begin{figure}[h]
  \includegraphics[width = \columnwidth ]{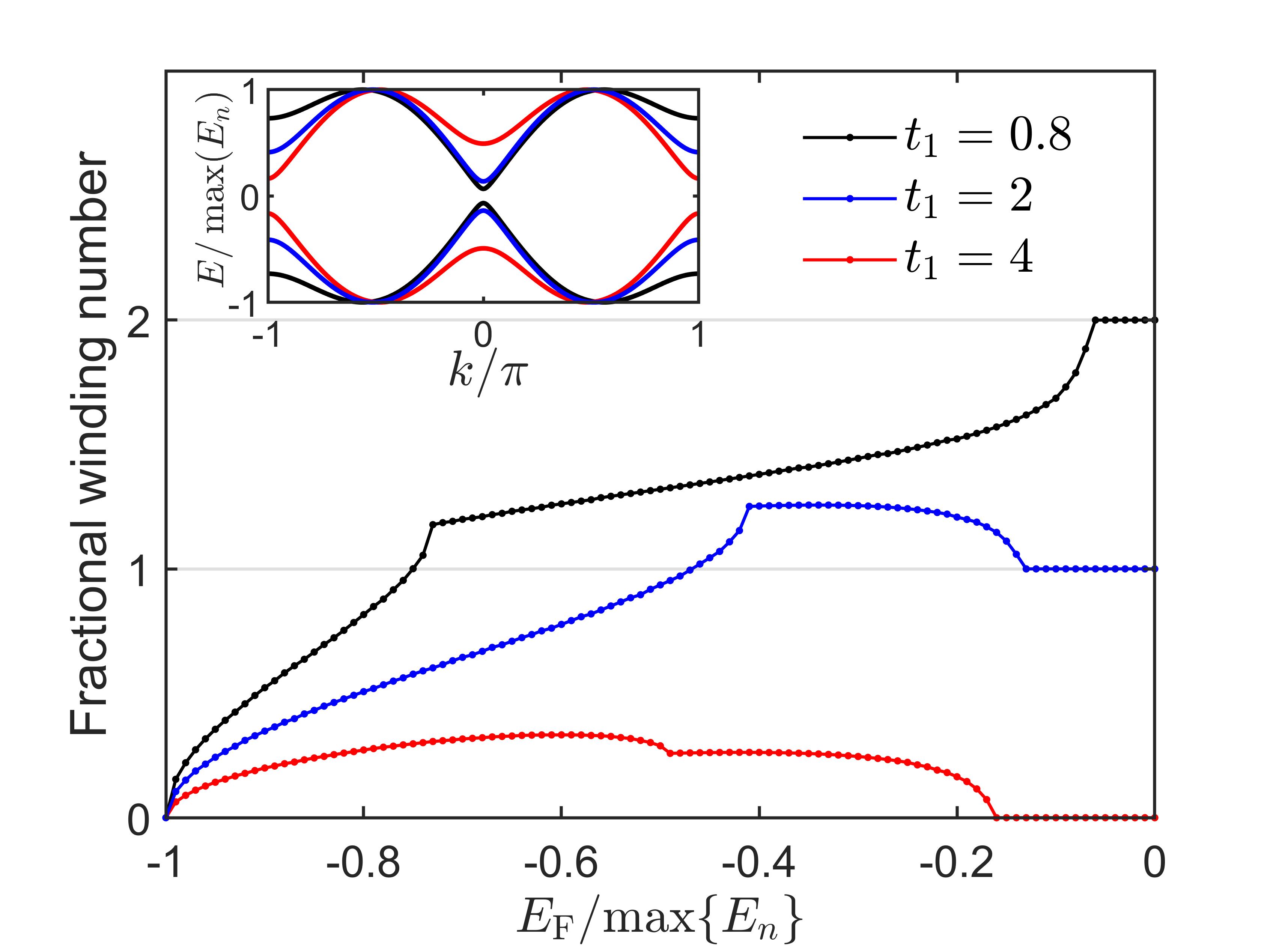}
  \caption{\label{fig:FIG_SSH_ScanEf}
 Winding number as a function of Fermi energy $E_F$ for three different parameters $t_1=0.8, 2, 4$ in the clean limit $W_1=W_2=W_3=0$.
 The inset shows the corresponding band structure.
 Other parameters are chosen as $t_2 = 1, t_3 = -2$.
 }
\end{figure}

Now, we calculate the winding number for various fillings (less than one-half) in the extended SSH model; see Fig.~\ref{fig:FIG_SSH_ScanEf}.
It can be seen that the winding number is zero when the filling is empty.
As the Fermi energy increases, the winding number changes continuously, and finally reaches an integer value when the system is half-filled (the Fermi energy $E_\mathrm{F}$ lies in the spectral gap).
This result may be understood by considering the winding number as the difference of polarization of $A$ and $B$ sublattices.
The difference in sublattice polarization will change with the filling numbers.
Notice that there appears discontinuity in the first derivative of the Fermi energy.
This is because there are some local minima in the band structure, as shown in the inset of Fig.~\ref{fig:FIG_SSH_ScanEf}.
This method can be also applied to the disordered case.

\section{Summary and Discussion}
\label{sec:discussion}

In summary, we propose a real-space formula for calculating the winding number of 1D chiral-symmetric systems.
Our real-space representation of the winding number is inspired by the projected position operator approach since the winding number can be written as the difference of polarization between two sublattices.
We show that our approach is equivalent to the momentum-space representation of the winding number in the clean limit.
Even in the presence of disorder, our formula produces a quantized value.
We have also shown that our method works for the case of fillings less than one-half.
With the help of TBC, we have proved the bulk-edge correspondence principle for the TBC winding number.
We further prove that our real-space representation of winding number is equivalent to the TBC winding number and the real-space formula proposed in Ref.~\citep{PhysRevLett.113.046802} in the thermodynamic limit at half filling.
Therefore our real-space representation of winding number also satisfies the bulk-edge correspondence and the self-averaging property.
Meanwhile, compared with the general TBC method, where includes the integral of twist angle $\Phi$, our formula can be obtained from a single Hamiltonian, which is more efficient.
Interestingly, we find that our real-space winding number can be expressed as a Bott index.
However, the Bott index is usually employed for the real-space Chern number \citep{Loring_2010, loring2014quantitative,toniolo2017equivalence}, which is quite different from the winding number in 1D chiral-symmetric topological insulator.
We also show that the Bott index has deep connection to the twisted boundary condition.
Our work may provide another concrete example for investigating the Bott index.
Meanwhile, one may find the position operator plays a crucial role in the construction of real-space representation of topological invariants, such as the Chern number \citep{PhysRevLett.105.115501,Prodan_2011,PhysRevB.84.241106,Loring_2010,yi2013coupling}, the Zak-Berry phase \citep{PhysRevB.26.4269,niu1991theory,RevModPhys.84.1419,PhysRevB.96.245115} and the winding number \citep{PhysRevLett.113.046802,PhysRevB.89.224203,PhysRevB.101.224201,PhysRevB.100.205302,PhysRevResearch.2.043233,meier2018observation,Maffei2018}.
Thus, it is intriguing to generalize the application of position operators to other topological systems in other topological classes or higher dimension in the future.
We also note that recently the SVD has been applied to non-Hermitian systems \citep{PhysRevA.99.052118}, where the singular values of a non-Hermitian Hamiltonian obey the bulk-edge correspondence.
The study on the interplay between non-Hermitian systems and disorder has received many interests recently \citep{PhysRevB.103.024205, claes2020skin}.
With the formula of winding number for the non-Hermitian system \citep{PhysRevX.8.031079}, it is intriguing to generalize our formula to the non-Hermitian case.
The non-Hermitian Hamiltonian can be considered as the off-diagonal block of some chiral-symmetric Hermitian Hamiltonians.
Then our real-space representation of the winding number may be extended to non-Hermitian systems.

\begin{acknowledgments}
This work has been supported by the National Natural Science Foundation of China (12025509,11874434), the Key-Area Research and Development Program of Guangdong Province (2019B030330001), and the Science and Technology Program of Guangzhou (201904020024). Y.K. is partially supported by the National Natural Science Foundation of China (Grant No. 11904419).
\end{acknowledgments}

\begin{figure}[!t]
	\includegraphics[width = \columnwidth ]{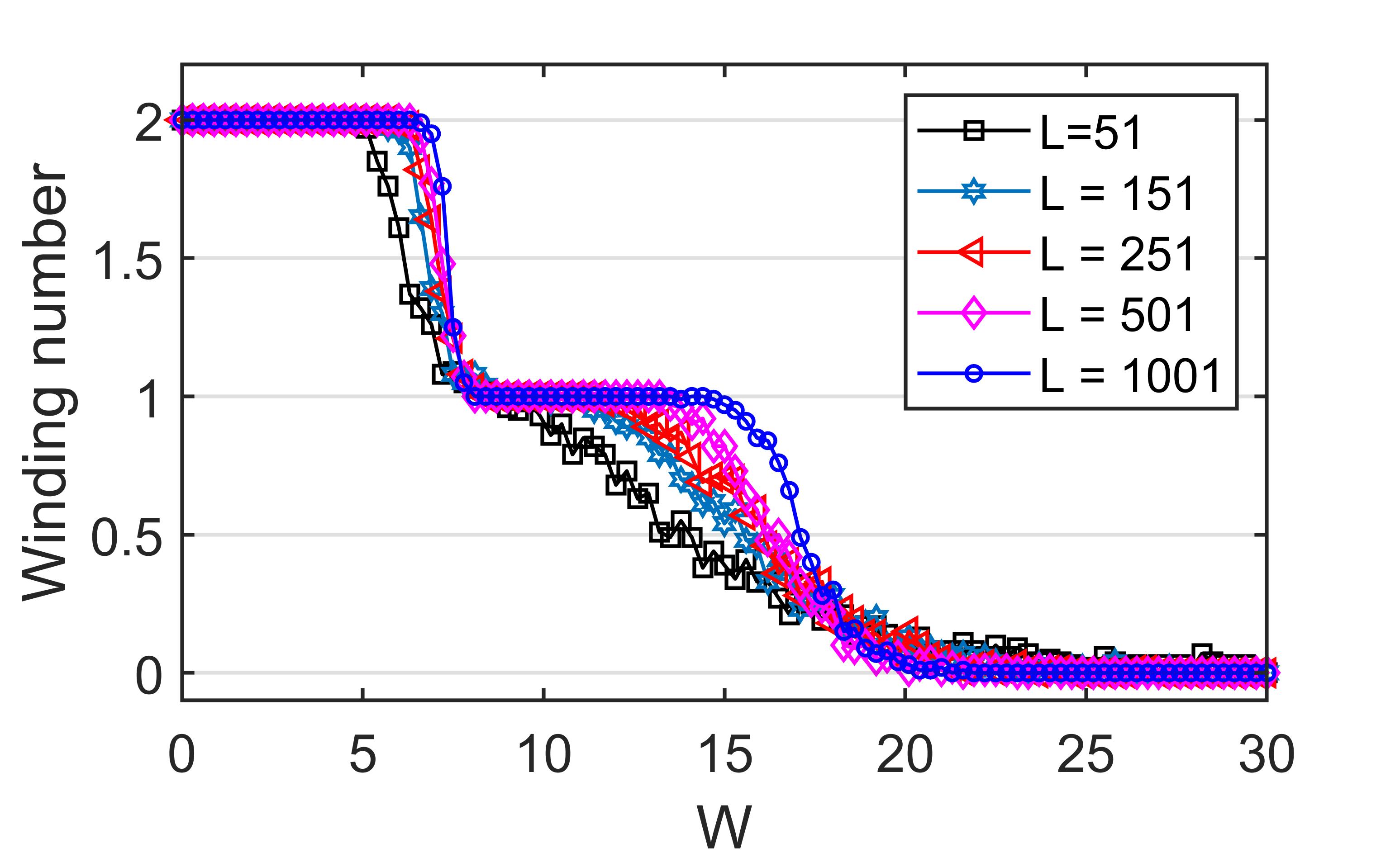}
	\caption{\label{fig:FIG_SSH_Scaling}
		Winding number as a function of disorder strength $W$ with different length $L$ of system.
		The results are averaged over 100 random realizations.
		Other parameters are the same as Fig.~\ref{fig:FIG_SSH_ScanW}.
	}
\end{figure}

\appendix
\section{Derivation of discretized formula of winding number}
\label{appendix:derivation_discrete_winding_number}
Here we give a detailed derivation of Eq.~\eqref{eqn:discretized_skew_polarization}.
Approximately, we have
\begin{equation}
U_\sigma ^{ - 1}\left( k \right){\partial _k}{U_\sigma }\left( k \right) \simeq U_\sigma ^{ - 1}\left( k \right)\frac{{{U_\sigma }\left( {k + \delta k} \right) - {U_\sigma }\left( {k} \right)}}{{\delta k}},
\end{equation}
which leads to
\begin{equation}
1 + U_\sigma ^{ - 1}\left( k \right){\partial _k}{U_\sigma }\left( k \right)\delta k \simeq U_\sigma ^{ - 1}\left( k \right){U_\sigma }\left( {k + \delta k} \right).
\end{equation}
Take the logarithm on both sides
\begin{eqnarray}
\log \left( {1 + U_\sigma ^{ - 1}\left( k \right){\partial _k}{U_\sigma }\left( k \right)\delta k} \right) \nonumber \\
\simeq \log \left( {U_\sigma ^{ - 1}\left( k \right){U_\sigma }\left( {k + \delta k} \right)} \right) ,
\end{eqnarray}
and use the approximation $\mathrm{log}(1+x)\approx x$ when $x\rightarrow 0$, we obtain
\begin{equation}
\label{eqn:appendix_final_approx}
U_\sigma ^{ - 1}\left( k \right){\partial _k}{U_\sigma }\left( k \right)\delta k \simeq \log \left( {U_\sigma ^{ - 1}\left( k \right){U_\sigma }\left( {k + \delta k} \right)} \right).
\end{equation}
In the thermodynamic limit, the quantities in two sides of Eq.~\eqref{eqn:appendix_final_approx} are equivalent.
Thus, the discretized form of winding number can be written as
\begin{equation}
\nu  = \frac{1}{{2\pi i}}\sum\limits_k {{\rm{Tr}}\left( {\log F_k^A - \log F_k^B} \right)},
\end{equation}
where $F_k^\sigma  = U_\sigma ^\dag \left( {k } \right){U_\sigma }\left( k - \delta k \right),\;\sigma  = A,B$.
Alternatively, we can use the fact that ${\rm{Tr}}\log \left( A \right) = \log \det \left( A \right)$ when $\left\| A \right\|<\pi$ \citep{aprahamian2014matrix}, and then obtain the expression in Eq.~\eqref{eqn:discretized_skew_polarization}.

\section{Scaling of the phase transition}
\label{appendix:scaling}
To examine the finite-size effect on the disorder-induced topological phase transition, we present the winding number as a function of disorder strength $W$ in Fig.~\ref{fig:FIG_SSH_Scaling}.
The results show that the boundary of phase transition tends to be clear when $L\to \infty$.

%
%
%

\bibliography{RealSpaceTopology_bib}

\end{document}